\documentclass[aps, prl, showpacs, superscriptaddress ,twocolumn ]{revtex4}
\usepackage{amsmath}
\usepackage{graphicx}
\usepackage{sidecap}

\begin{document}

\title{Bulk electronic structure of superconducting LaRu$_2$P$_2$ single crystals
    measured by soft x-ray angle-resolved photoemission spectroscopy}

\author{E. Razzoli}
\affiliation{Swiss Light Source, Paul Scherrer Institute, CH-5232 Villigen PSI, Switzerland}

\author{M.  Kobayashi }
\affiliation{Swiss Light Source, Paul Scherrer Institute, CH-5232 Villigen PSI, Switzerland}
\affiliation{Department of Applied Chemistry, School of Engineering, University of Tokyo, 7-3-1 Hongo, Bunkyo-ku, Tokyo 113-8656, Japan}

\author{V. N. Strocov}
\affiliation{Swiss Light Source, Paul Scherrer Institute, CH-5232 Villigen PSI, Switzerland}

\author{B. Delley}
\affiliation{Paul Scherrer Institute, CH-5232 Villigen PSI, Switzerland}

\author{Z. Bukowski}

\affiliation{Laboratory for Solid State Physics, ETH Z\"urich, CH-8093 Z\"urich, Switzerland}

\author{J. Karpinski}
\affiliation{Laboratory for Solid State Physics, ETH Z\"urich, CH-8093 Z\"urich, Switzerland}

\author{N. C. Plumb}
\affiliation{Swiss Light Source, Paul Scherrer Institute, CH-5232 Villigen PSI, Switzerland}

\author{M. Radovic}
\affiliation{Swiss Light Source, Paul Scherrer Institute, CH-5232 Villigen PSI, Switzerland}
\affiliation{Institut de la Matiere Complexe, EPF Lausanne, CH-1015, Lausanne, Switzerland}

\author{J. Chang}
\affiliation{Swiss Light Source, Paul Scherrer Institute, CH-5232 Villigen PSI, Switzerland}
\affiliation{Institut de la Matiere Complexe, EPF Lausanne, CH-1015, Lausanne, Switzerland}

\author{T. Schmitt}
\affiliation{Swiss Light Source, Paul Scherrer Institute, CH-5232 Villigen PSI, Switzerland}

\author{L. Patthey}
\affiliation{Swiss Light Source, Paul Scherrer Institute, CH-5232 Villigen PSI, Switzerland}

\author{J. Mesot}
\affiliation{Swiss Light Source, Paul Scherrer Institute, CH-5232 Villigen PSI, Switzerland}
\affiliation{Institut de la Matiere Complexe, EPF Lausanne, CH-1015, Lausanne, Switzerland}

\author{M. Shi}
\affiliation{Swiss Light Source, Paul Scherrer Institute, CH-5232 Villigen PSI, Switzerland}

\begin{abstract}

We present a soft X-ray angle-resolved photoemission spectroscopy (SX-ARPES) study of the stoichiometric pnictide superconductor
 LaRu$_2$P$_2$. The observed electronic structure is in good agreement with density functional theory (DFT) calculations. 
However, it is significantly different from its counterpart in high-temperature superconducting Fe-pnictides. In particular the 
bandwidth renormalization present in the Fe-pnictides ($\sim$ 2 - 3) is negligible in LaRu$_2$P$_2$ even though the mass 
enhancement is similar in both systems. 
Our results suggest that the superconductivity in LaRu$_2$P$_2$ has a different origin with respect to the iron
pnictides. Finally we demonstrate that the increased probing depth of SX-ARPES, compared to the widely used ultraviolet ARPES, is essential in determining the bulk electronic structure in the experiment.

\end{abstract}

\date{\today}
\maketitle

Before the discovery of high-temperature Fe-pnictide superconductors
in 2008~\cite{Kamihara}, superconductivity had been observed in
a number of stoichiometric pnictides, with LaRu$_2$P$_2$ having the
highest superconducting transition temperature (T$_c$) of $\sim$ 4
K~\cite{Jeitschko}. For Fe-pnictides at ambient pressure, most of
the undoped compounds (e.g. BaFe$_2$As$_2$) have an
antiferromagnetic (AFM) ground state, and the superconductivity
emerges by applying external pressure or through chemical
substitution(s). For example, partially substituting Ru for Fe or P for As can
turn BaFe$_2$As$_2$ into superconductors~\cite{Kasahara, Sharma}. On the
 other hand, the fully substituted compounds LaFe$_2$P$_2$~\cite{Jeitschko2} and 
 $M$Ru$_2$P$_2$ ($M$ = Ca, Sr, Ba) are nonsuperconducting down
  to $1.8$ K~\cite{Jeitschko}. Low-temperature superconductivity occurs when
 Fe is replaced by Ru in LaFe$_2$P$_2$ or the alkaline-earth atom is
  replaced by La in $M$Ru$_2$P$_2$. This prompts two
intriguing questions: What are the differences in their electronic
structures that cause Fe-pnictide superconductors to have
significantly higher T$_c$'s than superconducting stoichiometric
pnictides?; and is the superconducting mechanism the same in both
systems?  Recent quantum oscillation measurements have indeed shown
similarities in the mass enhancement in Fe-pnictides and
LaRu$_2$P$_2$~\cite{Moll}. However, this technique is intrinsically
unable to distinguish between correlations due to electron-electron
interactions, which usually renormalize the entire bandwidth, and
contributions due to electron-bosonic mode coupling, which is
limited to a narrow energy range (of the order of the boson energy)
around E$_F$. It would be particularly interesting to compare the
Fermi surface (FS) topologies and electron-electron correlation
effects in LaRu$_2$P$_2$ and Fe-pnictides, since these have been
thought to play important roles in the emergence of
superconductivity in Fe-pnictides~\cite{Paglione}.

To address this question we probe the electronic structure of
stoichiometric superconducting pnictide LaRu$_2$P$_2$ using
soft X-ray angle-resolved photoemission spectroscopy (SX-ARPES).
LaRu$_2$P$_2$ is isostructural with ``122''  Fe-pnictide
superconductors (e.g. doped BaFe$_2$As$_2$), whose electronic states
so far have been mostly studied by various conventional techniques~\cite{Paglione}.
We show that both the FS topology and the bandwidth renormalization
of LaRu$_2$P$_2$ are very different from their high-temperature
superconducting Fe-pnictide counterparts. We also demonstrate that,
compared to ARPES using ultraviolet light as the excitation source
(UV-ARPES), the increased photoelectron mean free path of SX-ARPES
is essential in determining of the bulk electronic structure of
pnictides. We highlight that it will be important to apply SX-ARPES
to Fe-pnictides to resolve the controversies in the reported
electronic structures obtained by different techniques.

The SX-ARPES experiments were performed at the Advanced Resonant Spectroscopies
(ADRESS) beamline at the Swiss Light Source (SLS). 
Data were collected using \textit{p}-polarized light with an  overall energy resolution on the order of 80 meV. The UV-ARPES experiments were carried out at the Surface and Interface Spectroscopy (SIS) beam line at
SLS.  The overall energy resolution was about 15 meV. The samples were
cleaved \textit{in situ} at 10 K  and measured in a vacuum always better than $5 \times 10^{-11}$ mbar.  For details on samples and experimental procedures, see supplemental material~\cite{Supplement}.

\begin{figure}[h!]
\includegraphics[width=0.48\textwidth]{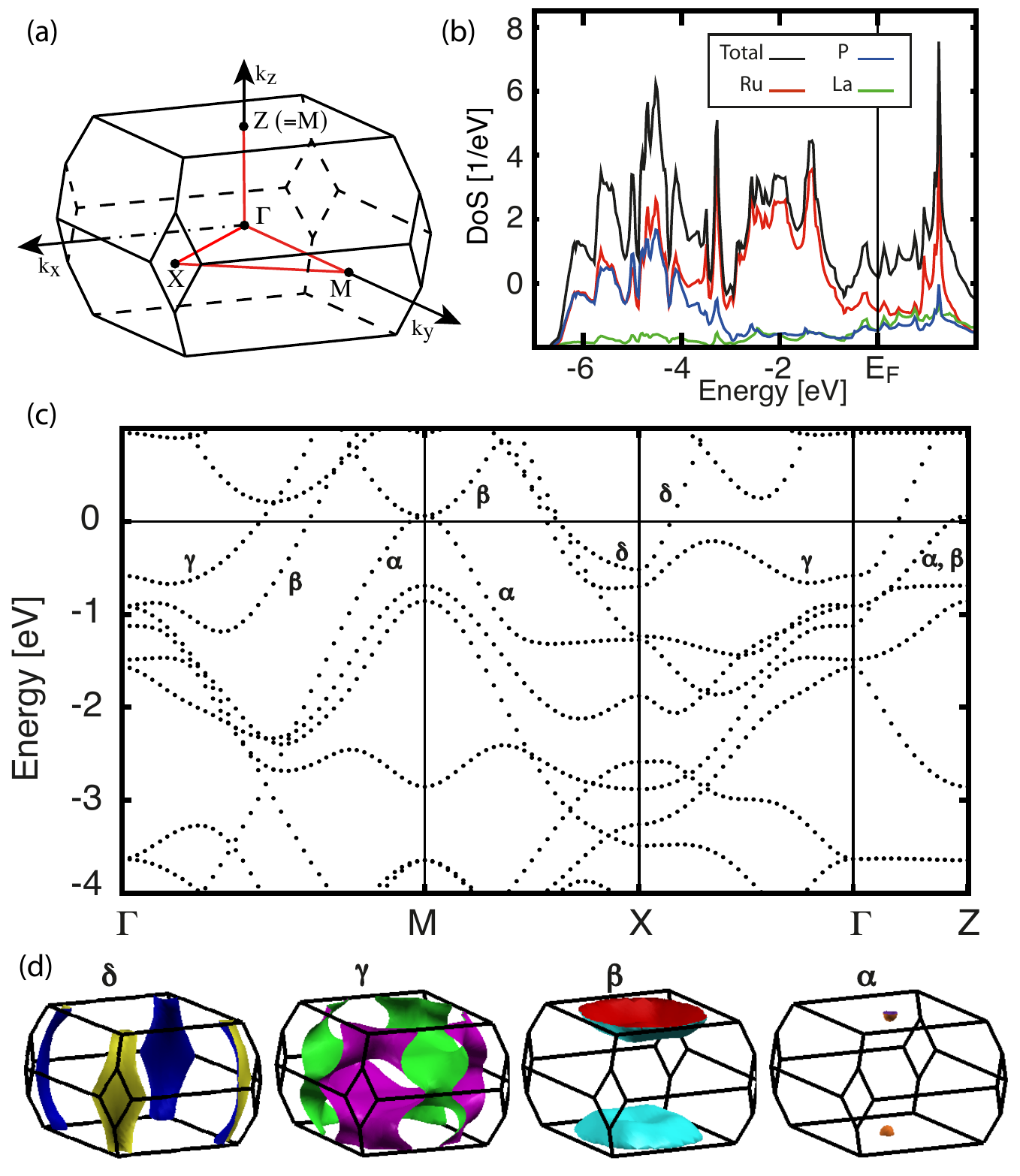}
\caption{(Color online) The electronic structure and FS of
LaRu$_2$P$_2$ from DFT calculation. (a) Brillouin zone (BZ)
 (X = ($\pi$/\textit{a}, $\pi$/\textit{a}, 0), M = (0, 2$\pi$/\textit{a}, 0) and Z = (0, 0, 2$\pi$/\textit{c})). (b)
Total and partial density of state (DOS) with different atomic
characters. (c) The band structure along high symmetry lines. (d)
Three-dimensional FS sheets of bands labeled $\alpha$,
$\beta$, $\gamma$ and $\delta$ in (c) plotted using XcrysDen visualization package \cite{XCrysden}.}
\label{fig_DFT}
\end{figure}

Figure~\ref{fig_DFT} shows the band structure in the Brillouin zone
(BZ) calculated from  DFT \cite{Hohenberg} using the DMol$^3$ method \cite{Delley}. Lattice
parameters (\textit{a} =  4.031 $\mathring{\text{A}}$, \textit{c} =
10.675 $\mathring{\text{A}}$) and internal atomic positions
(\textit{z}$_{\text{La}}$=0, \textit{z}$_{\text{Ru}}$= 0.25,
\textit{z}$_\text{P}$= 0.3593) determined from X-ray diffraction
\cite{Jeitschko} are used in the calculation. DFT total-energy
minimization resulted in \textit{z}$_P$= 0.36165, which gave only
minor changes in the band structure. Near the Fermi level ($E_F$),
$\alpha$, $\beta$ and $\delta$ have predominant Ru-4\textit{d}
characters (Fig.~\ref{fig_DFT}(b)), analogous to the case of
Ba$_{1-x}$K$_{x}$Fe$_2$As$_2$ for which the bands within 2 eV of
$E_F$ are formed mainly by Fe-3\textit{d} electrons \cite{Singh}.
The only exception is the $\gamma$ band, which has admixed
Ru-4\textit{d} and La orbital character. All bands that cross $E_F$
disperse strongly in the $k_z$ direction, hence their FS sheets are
highly three-dimensional (Fig.~\ref{fig_DFT}(c)-(d)).

\begin{figure}[h!]
\includegraphics[width=0.49\textwidth]{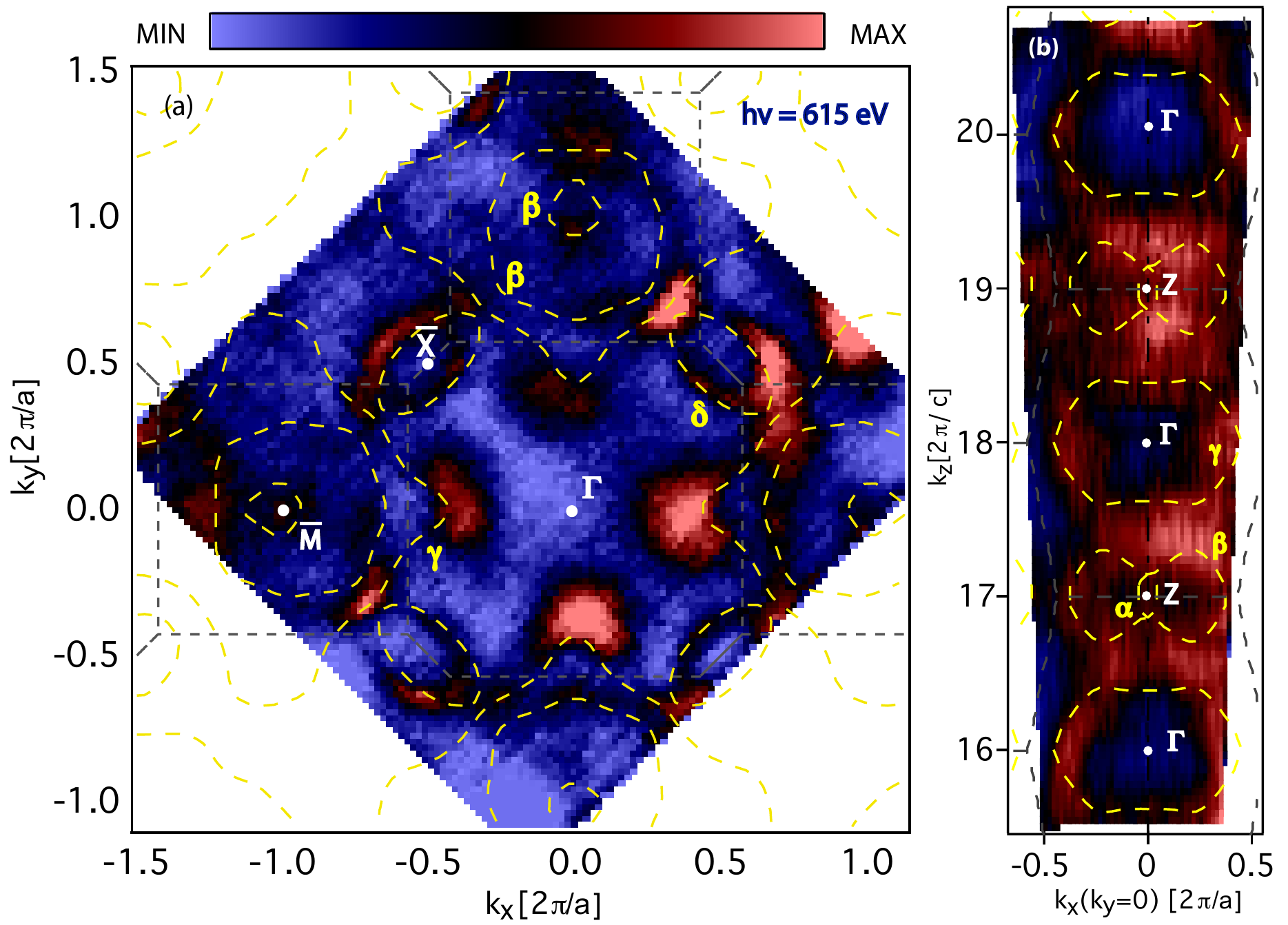}
\caption{(Color online) FS intensity maps obtained  with SX-ARPES.
(a) FS map near the ($k_x$,$k_y$,$22\times 2\pi/c$) plane,
taken with $h\nu = 615$ eV. (b) FS map in the ($k_x$,0,$k_z$) plane,
taken with $h\nu = 300 - 550 $ eV in steps of 5 eV. The maps in (a)
and (b) are obtained by integrating ARPES spectral weight in an
energy window of $E_F\pm 50$ meV. The superimposed dashed lines are
the FS from DFT calculation. $\bar{M}$ ($\bar{X}$) is close to, but
slightly different from, M (X) in k$_z$ value.} \label{FSsMaps}
\end{figure}

In Fig.~\ref{FSsMaps}(a)-(b) we plot the ARPES spectral weight
mapping at $E_F$ near  ($k_x$,$k_y$,$22\times 2\pi/c$) and in the
($k_x$,0,$k_z$) planes. The $k_z$ values were
extracted by using the free-electron final-state
approximation~\cite{Hufner} with an inner potential ($V_0 = 15$ eV)
estimated by observing the periodicity of the FS as a
function of $h\nu$ and taking into account the photon momentum for
our experimental geometry. The superimposed dashed lines are the FS
from the DFT calculation. It should be noted that the calculated FS
in Fig.~\ref{FSsMaps}(a)  takes into account the change in $k_z$ as
a function of the emission angles of the photoelectrons.

\begin{figure*}
\includegraphics[width=1\textwidth]{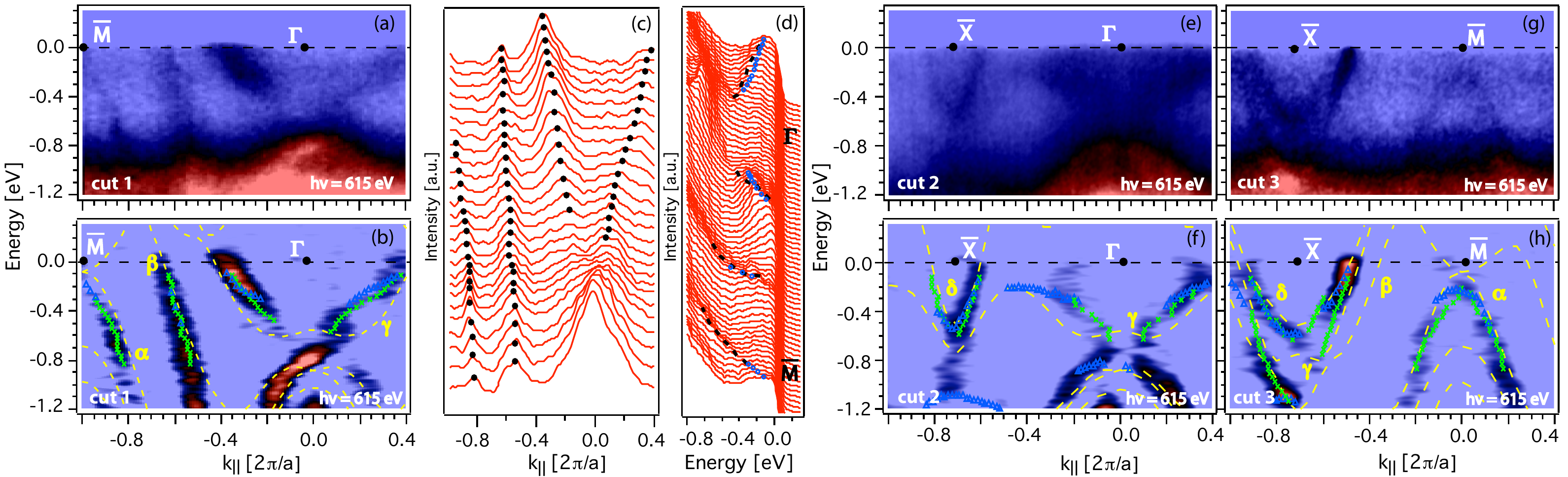}
\caption{(Color online) Band dispersion in  LaRu$_2$P$_2$ measured
with  SX-ARPES.  (a), (e) and (g) ARPES spectrum along
M-$\Gamma$ direction (cut 1 Fig.~\ref{SurfaceBands} (b)), $\Gamma$-X
direction (cut 2 Fig.~\ref{SurfaceBands} (b))
 and the M-X  direction (cut 3 Fig.~\ref{SurfaceBands} (b)),
respectively ($h\nu= 615$ eV).  (b),(f) and (h) Second derivatives
of the MDCs in (a),  (e) and (g),
respectively. The DFT band structure (dashed lines), MDC peaks (green crosses) and EDC peaks (blue triangles) are superimposed on the data.
 (c) MDCs and MDC peaks (black dots) from (a). (d) EDCs and EDC peaks (blue circles) from (a). Dashed lines trace the MDCs dispersion.}   \label{BandDisp}
\end{figure*}

\begin{figure}[h!]
\includegraphics[width=0.48\textwidth]{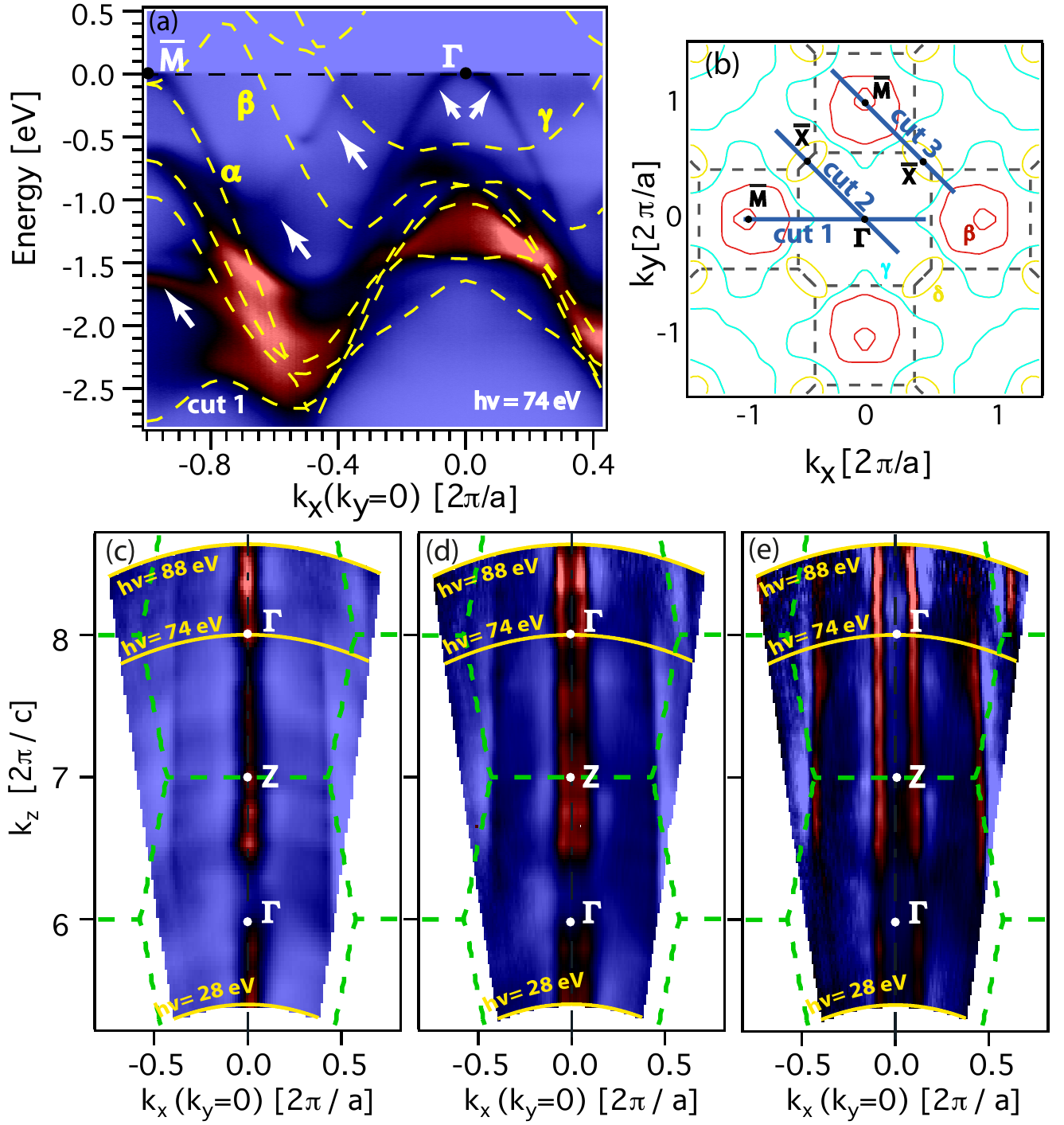}
\caption{(Color online) Band dispersion in  LaRu$_2$P$_2$ measured
with UV-ARPES. (a) UV-ARPES spectrum along the M-$\Gamma$
direction (cut 1 in (b)) taken with $h\nu = 74$ eV. The superimposed
dashed lines are the band structure from the DFT calculation. (b) DFT
Fermi surface for $h\nu= 615$ eV. (c)-(e) Intensity map in
($k_x$,0,$k_z$) plane, taken with $h\nu = 28-88 $ eV in steps of 2
eV. The maps are obtained by integrating ARPES spectral weight in an
energy window of $E_F\pm 10$ meV, $(E_F -100$ meV$ )\pm 10$meV and
$(E_F-200$ meV$) \pm 10$ meV, respectively.}\label{SurfaceBands}
\end{figure}

Figure~\ref{BandDisp}(a) shows the ARPES intensity as a function of
energy and $k$ along cut 1 in Fig.~\ref{SurfaceBands}(b). This cut
is close to, but slightly deviates from, the  $\Gamma$-M axis because the
$k_z$ values depend on the emitting angles of photoelectrons for a
fixed $h\nu$. To trace the detailed dispersion of the $\alpha$,
$\beta$ and $\gamma$ bands up to 1.2 eV below $E_F$, we present in
Fig.~\ref{BandDisp}(b), (c) and (d) the second derivative intensity plot along
with the bands from the DFT calculation,  the momentum distribution curves (MDCs) and the energy distribution curves (EDCs),
 respectively. The overall agreement
between the ARPES spectra and the calculated electronic structure is
significant; all features in the ARPES data can be identified from
the DFT calculation. A quantitative agreement can be achieved after
shifting the $\gamma$ band up by 100 meV and shifting other bands
down slightly for maintaining the charge neutrality of the system.
It should be mentioned that, in the
study of the electronic structure of SrFe$_2$P$_2$, shifts of
the calculated electronic structure were also required to accurately
reproduce the FS from quantum oscillations
measurements~\cite{Analytis}. The shifts were thought of as a fine
tuning of the band structure calculations.

In contrast to Fe-pnictide superconductors, whose bandwidths are
renormalized by a factor of 2-3~\cite{Terashima,Richard}, the
dispersions in Fig.~\ref{BandDisp} can be reproduced by DFT
calculation without notable renormalization. To investigate the
bandwidth renormalization effect further we have measured the band
dispersion along various symmetry lines, as well as many
off-symmetry lines in the BZ. Figure~\ref{BandDisp} (e)-(h) shows
the ARPES spectra acquired along the $\Gamma$-X and M-X directions
(cuts 2 and 3 in Figure~\ref{SurfaceBands}(b)) and their second
derivative intensity plots. EDC and MDC peak positions are superimposed on the second derivative intensity plots. As for the $\alpha$, $\beta$ and
$\gamma$ bands, the dispersion of the $\delta$ band which forms the
electron-like FS pocket around the X point can also be well
described by the DFT calculation. The good agreement between the
electronic structure determined experimentally and that calculated
from DFT indicates that the Ru-4\textit{d} electrons are delocalized
in nature, which can be captured well by DFT band calculations.

The negligible bandwidth renormalization indicates that the overall
electron-electron correlation is weaker in LaRu$_2$P$_2$ than in BaFe$_2$As$_2$.  
Such a reduction is expected when the ratio of the Coulomb repulsion $U$ to the DFT bandwidth $W_{DFT}$ is decreased~\cite{Fazekas}.
The calculated $U$ is similar in Fe
3\textit{d} and Ru 4\textit{d}~\cite{Ersoy}, while our DFT calculations show that the bandwidth of LaRu$_2$P$_2$ ($\sim$~10 eV) is substantially
larger than that of BaFe$_2$As$_2$ ($\sim$~6 eV). This is not surprising since 4\textit{d} orbitals are
much more extended than their 3\textit{d} counterparts due to higher principal quantum number.
We thus conclude that the weaker correlation
strength can be attributed to the larger bandwidth  of the bare
DFT bands of LaRu$_2$P$_2$. 
Additionally, the increase in the band filling $n$ due to the replacing Ba with La may further reduce the electron-electron correlations~\cite{Ishida}.
The observed bandwidth renormalization is consistent with the trend of weakening correlation strength in Ba(Fe$_{1-x}$Ru$_x$)$_2$As$_2$ and
BaFe$_2$(As$_{1-x}$P$_x$)$_2$ with increasing $x$ , which are isoelectric substitutions~\cite{Brouet,Shishido}. 
The general feature of these substitutions, as well as the substitution
of Sr with La in Sr$_{1-x}$La$_x$Fe$_2$As$_2$~\cite{Muraba}, is that
at low temperatures upon increasing the concentration of Ru (P, La),
superconductivity emerges from an AFM spin-density-wave state, and
further substitution drives the system towards a paramagnetic
nonsuperconducting metal. The superconductivity occurs only in a
finite range of substitution, and the fully substituted compounds
(BaRu$_2$As$_2$, BaFe$_2$P$_2$) are nonsuperconducting. This
suggests that the low-temperature superconductor, LaRu$_2$P$_2$,
belongs to a superconducting phase which does not connect to the
high-temperature superconducting phase located near the AFM region of the temperature-substitution phase diagram.
Instead this superconducting
phase is ``conventional'' in the sense that superconductivity
emerges from a Fermi liquid-like normal state.
Since our study shows that the bandwidth renormalization of
LaRu$_2$P$_2$ is negligible,  a natural explanation for the mass enhancement observed in
quantum oscillation experiments~\cite{Moll} is that the renormalization 
occurs only near E$_F$ and is caused by an electron-bosonic mode interaction with the mode energy $\lesssim$ 80 meV, which is smaller than the energy resolution used in our SX-ARPES measurements.
This is different to Fe-pnictides whose mass enhancement (by a factor of
2-3) results from bandwidth renormalizations and is attributed to electron-electron interaction~\cite{Supplement,Terashima,Richard}.

We now emphasize that the increased probing depth in SX-ARPES,
compared to widely used UV-ARPES, is critical in revealing the bulk
electronic structure of LaRu$_2$P$_2$. For comparison we plot in
Fig.~\ref{SurfaceBands}(a) a UV-ARPES spectrum along the $\Gamma$-M
direction acquired using circularly polarized light   with $h\nu$ = 74 eV (data
measured with \textit{s}- and \textit{p}-polarized light show similar behavior). The bands indicated by
arrows, which do not appear in SX-ARPES spectra (Fig.~\ref{BandDisp}
(a)-(b)), cannot be identified with the bulk electronic structure
from the DFT calculation. The dispersion of these bands does not change
with photon energies in the range of 28 - 88~eV which corresponds to
a $k_z$ range covering more than one BZ
(Fig.~\ref{SurfaceBands}(c)-(e)). This suggests that these new
bands originate from the surface layer and do not represent the bulk
electronic structure.  Since LEED shows a clear $(1\times 1)$ pattern~\cite{Supplement}, we conclude that  the additional bands observed with UV-ARPES may have the same origin 
as those in the ``1111'' Fe-pnictides~\cite{Eschrig}, where they result from the surface and subsurface layers due to surface relaxation.
  It is interesting to note that some FS pockets of BaFe$_2$As$_2$ observed
by UV-ARPES are absent in the FS obtained from bulk-sensitive
Shubnikov–-de Haas oscillation measurements~\cite{Terashima}. 
Moreover it has been reported that the energy bands obtained from UV-ARPES on
BaFe$_{2-x}$Co$_x$As$_2$ are significantly distorted with respect to
the bulk electronic structure~\cite{Heumen}. It will be important to
re-visit Fe-pnictides with SX-ARPES to clarify this controversy. It
will be vital for the community to gain confidence in ARPES results
on the bulk electronic structure of these materials, especially for
those studies for which there is no independent experimental check
on bulk electronic structure, such as quantum oscillation
experiments, for example on type-II superconductors with large
$H_{c2}$ values.

In summary, using SX-ARPES we revealed the electronic structure in
the normal state of the ``122'' Ru-pnictide superconductor
(LaRu$_2$P$_2$), which is in significant agreement with DFT
calculations. The negligible renormalization of the bandwidth suggests that the mass enhancement observed in
quantum oscillation experiments is due to electron-boson coupling and is limited to narrow range ($\lesssim 80$ meV) around the E$_F$.
Our results suggest that the origin of the superconducting phase in
LaRu$_2$P$_2$ is different from the one in the ``122'' Fe-pnictides
and more ``conventional'' in the sense that it emerges from a Fermi
liquid-like normal state. By comparing with UV-ARPES spectra, we
have illustrated that the increased probing depth of SX-ARPES is
essential for the determination of bulk electronic structure in our
experiments.

We are grateful to P. J. W. Moll and B. Batlogg for useful discussions.
M.K. acknowledges support from the Japan Society for the Promotion of Science.
This work was supported by the Swiss National Science Foundation through NCCR MaNEP.
This work was performed at SLS of the Paul Scherrer Institut, Villigen PSI, Switzerland.
We thank the beam line staff of ADRESS and of SIS for their excellent support.



\end{document}